\documentclass[pdflatex,sn-mathphys-num]{sn-jnl}% Math and Physical Sciences 

\makeatletter
\def\abstractfont{%
  \reset@font
  \fontsize{9bp}{11bp}
  \selectfont
  \unboldmath
  \leftskip=24pt
  \rightskip=24pt
  \parfillskip=0pt plus 1fil
}
\makeatother
\usepackage{graphicx}%
\usepackage{multirow}%
\usepackage{amsmath,amssymb,amsfonts}%
\usepackage{amsthm}%
\usepackage{mathrsfs}%
\usepackage[title]{appendix}%
\usepackage{xcolor}%
\usepackage{textcomp}%
\usepackage{manyfoot}%
\usepackage{booktabs}%
\usepackage{algorithm}%
\usepackage{algorithmicx}%
\usepackage{algpseudocode}%
\usepackage{listings}%
\usepackage[version=4]{mhchem}

\theoremstyle{thmstyleone}%
\theoremstyle{thmstyletwo}%

\theoremstyle{thmstylethree}%

\begin{document}

\title[Article Title]{\textbf{Exchange splitting as a descriptor for giant anomalous Hall and Nernst effects in ferromagnets}}

\author*[1]{\fnm{Ivan} \sur{Kurniawan}}\email{kurniawan.ivan@nims.go.jp}

\author[2]{\fnm{Guangzong} \sur{Xing}}

\author[1,3]{\fnm{Yoshio} \sur{Miura}}

\author[1]{\fnm{Keisuke} \sur{Masuda}}

\affil*[1]{National Institute for Materials Science (NIMS), Tsukuba 305-0047, Japan}

\affil[2]{Zhejiang Key Laboratory of Magnetic Materials and Applications, Ningbo Institute of Materials Technology and Engineering, Chinese Academy of Sciences, Ningbo 315201, China}

\affil[3]{Kyoto Institute of Technology, Kyoto, 606-8585, Japan}

\abstract{The anomalous Hall effect (AHE) and anomalous Nernst effect (ANE), which describe transverse electrical and thermoelectric responses in magnetic materials, respectively, are promising for spintronic and energy-harvesting applications. Here, we employ high-throughput first-principles calculations to investigate 2251 chemically substituted tetragonal $L1_0$ alloys. Among ferromagnets, enhanced responses emerge preferentially in alloys derived from parent compounds with small exchange splitting: no alloy derived from FePt, the archetypal $L1_0$ ferromagnet, reaches the high-response regime, whereas NiPt- and CoIr-derived alloys occupy it in large numbers. Small exchange splitting keeps majority- and minority-spin bands near the Fermi level, giving chemical substitution more opportunity to modify near-Fermi-level band crossings and amplify the Berry curvature. We predict a giant anomalous Hall conductivity of $2809\,\mathrm{S\,cm^{-1}}$ in \ce{(Co_{0.8}Fe_{0.2})(Ir_{0.7}Pt_{0.3})} and a giant anomalous Nernst conductivity of $7.72\,\mathrm{A\,m^{-1}\,K^{-1}}$ in \ce{(Ni_{0.8}Co_{0.2})(Pt_{0.7}Ir_{0.3})}. Our results identify the exchange splitting of the parent compound as a descriptor for chemical tunability toward giant Berry-curvature-driven transport responses.}

\keywords{anomalous Hall effect, anomalous Nernst effect, parent-compound, chemical substitution, exchange splitting}

\maketitle

\section{Introduction}\label{sec1}

Transverse charge and thermoelectric transport in magnetic materials, known as the anomalous Hall effect (AHE) and anomalous Nernst effect (ANE), have attracted considerable interest because of their potential applications in spintronic and energy-harvesting technologies. In conventional ferromagnets, the magnitudes of these responses were historically discussed in relation to magnetization. However, the observation of large anomalous transport responses in topological ferromagnets and noncollinear antiferromagnets has demonstrated that the intrinsic AHE and ANE are governed by the Berry curvature of the electronic bands rather than by the net magnetization alone \cite{nagaosa2010,guin2019,sumida2020,nakatsuji2015,ikhlas2017,sakai2018,liu2018,sakai2020}. Electronic-structure features near the Fermi level therefore play a central role in realizing large anomalous Hall conductivity (AHC) and anomalous Nernst conductivity (ANC). Despite the discovery of large AHC and ANC in diverse magnetic systems, broadly applicable materials-design principles remain limited.

High-throughput computational and experimental approaches provide powerful routes for exploring chemically substituted materials spaces. Nevertheless, the number of possible compositions increases rapidly with the number of constituent elements and substitution ratios, making exhaustive exploration costly in terms of both computational and experimental resources. High-throughput screening alone may therefore remain inefficient unless it is complemented by a design principle that identifies promising parent compounds before extensive compositional optimization. An important question is thus which electronic characteristics make a parent compound highly responsive to chemical substitution and capable of yielding derivatives with giant AHE and ANE.

In ferromagnets, exchange splitting determines the relative energy positions of majority- and minority-spin bands. A relatively small exchange splitting can leave more bands from both spin channels near the Fermi level, thereby increasing the opportunities for chemical substitution to modify band crossings and other electronic-structure features associated with large Berry curvature. If this tendency can be established systematically, exchange splitting may serve as a practical descriptor for selecting parent compounds with high chemical tunability toward large AHC and ANC. Such a descriptor would not only improve the efficiency of computational materials screening but could also guide combinatorial synthesis and high-throughput experimental characterization by narrowing the search to promising compositional families.

Tetragonal $L1_0$-ordered alloys provide a well-suited platform for investigating how chemical substitution modifies anomalous transport and why certain parent compounds are more tunable toward giant AHE and ANE. Their relatively simple crystal structure, robust ferromagnetism, substantial spin--orbit coupling, and strong magnetic anisotropy make them promising candidates for applications such as thermoelectric waste-heat harvesting \cite{uchida2021} and AHE-based magnetic-recording read heads \cite{nakatani2024}. Canonical systems such as FePt and related $L1_0$ alloys have been extensively studied because of their attractive magnetic and transport properties \cite{weller2000,okamoto2002,mizukami2011,kurniawan2023,seemann2010,he2012,mizuguchi2012,shi2020}. Composition-spread thin-film studies of these systems have further shown that the largest anomalous transport responses need not occur at stoichiometry: in \ce{Co_{1-x}Pt_x}, the largest ANC is obtained for a Pt-rich $L1_0$ composition rather than the nearly stoichiometric one \cite{toyama2024}, while in \ce{Fe_{1-x}Pt_x} the largest AHC appears at an Fe-rich composition associated with the $L1_2$-\ce{Fe3Pt} phase \cite{patel2025}. Meanwhile, previous high-throughput and systematic investigations of AHC and ANC have focused mainly on stoichiometric compounds across broader material families \cite{noky2020,tanzim2023,zelezny2023}. Consequently, chemically substituted $L1_0$ alloys remain insufficiently explored, and the roles of parent-compound choice and compositional tuning in Berry-curvature-driven transport are not yet well understood.

In this work, we perform a high-throughput first-principles investigation of chemically substituted $L1_0$ alloys using the workflow previously implemented for cubic Heusler systems by some of the present coauthors~\cite{xing2024}. We show that parent-compound choice plays a decisive role in determining the tunability of AHC and ANC upon chemical substitution. In particular, within the ferromagnetic subset, alloys derived from parent compounds with relatively small exchange splitting tend to retain more bands from both spin channels near the Fermi level, creating favorable opportunities for chemical substitution to generate electronic structures associated with large Berry curvature. Our results identify reduced exchange splitting as a useful predictive descriptor for selecting ferromagnetic parent compounds that can yield large anomalous Hall and Nernst responses upon chemical substitution.

\section{Results}\label{sec2}
\subsection{Distribution of AHC and ANC across compositions}\label{subsec2}

\begin{figure}[p] % Do NOT use \begin{figure*}
	\centering
	\includegraphics[width=\textwidth]{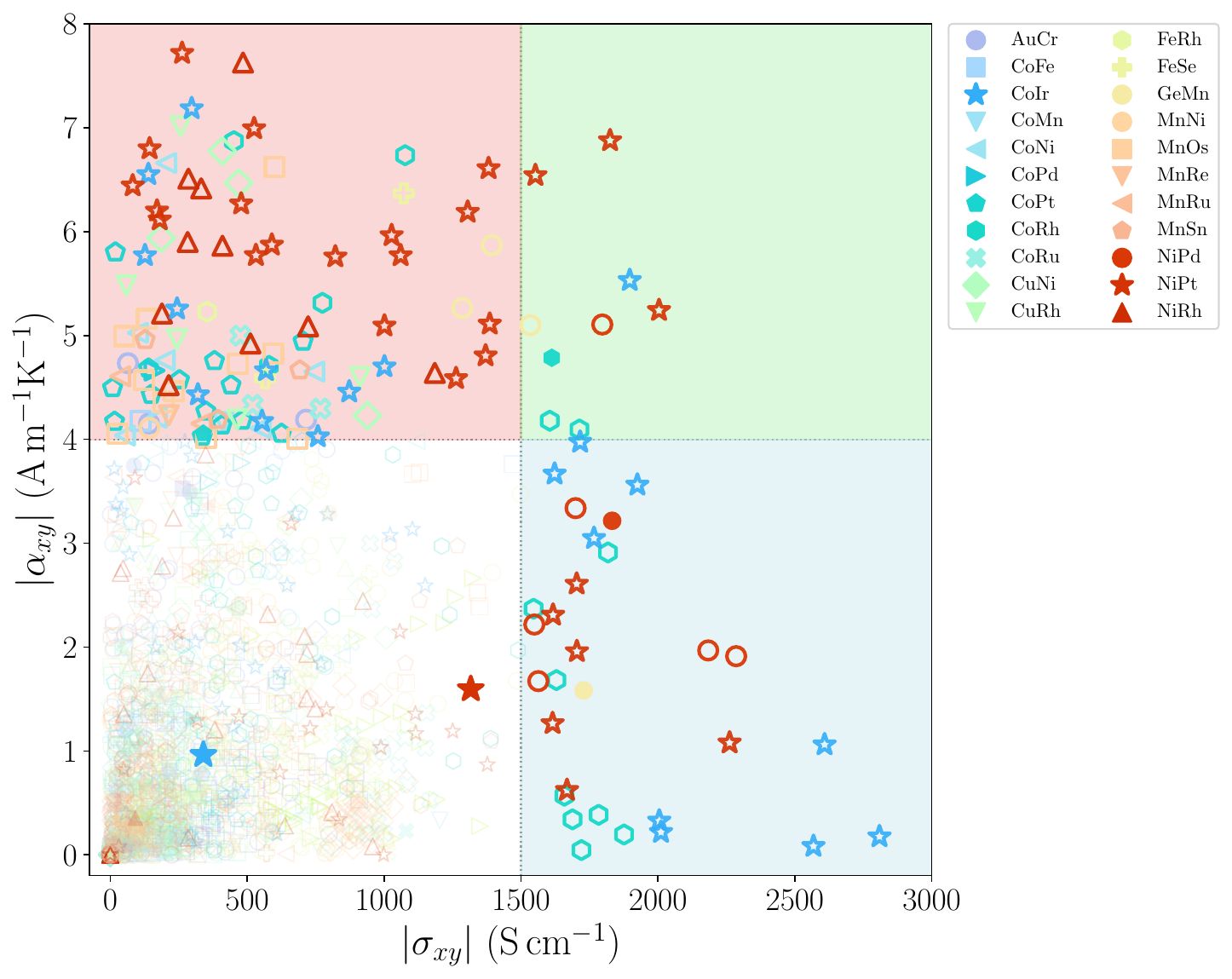} % for an image file named example_figure.*
\caption{\textbf{Distribution of $|\sigma_{xy}|$ and $|\alpha_{xy}|$.}
Absolute values of $\sigma_{xy}$ at $T=0\,\mathrm{K}$ and $\alpha_{xy}$ at $T=300\,\mathrm{K}$ for the 2251 investigated $L1_0$ alloy compositions. Filled symbols represent stoichiometric parent compounds, whereas open symbols of the same color represent chemically substituted compounds derived from the corresponding $XY$ parent system. The red and blue regions indicate candidates satisfying $|\alpha_{xy}| \ge 4\,\mathrm{A\,m^{-1}\,K^{-1}}$ and $|\sigma_{xy}| \ge 1500\,\mathrm{S\,cm^{-1}}$, respectively, while the green region contains candidates satisfying both criteria.
}
	\label{fig1} % give each figure a logical label name
\end{figure}

Figure~\ref{fig1} summarizes the absolute values of the anomalous Hall conductivity ($\sigma_{xy}$) and anomalous Nernst conductivity ($\alpha_{xy}$) calculated for 2251 $L1_0$ alloy compositions. Rather than being uniformly distributed across chemical space, large transverse responses cluster around alloys derived from a limited subset of parent compounds. In particular, NiPt- and CoIr-based systems dominate the region defined by $|\sigma_{xy}| \ge 1500\,\mathrm{S\,cm^{-1}}$ or $|\alpha_{xy}| \ge 4\,\mathrm{A\,m^{-1}\,K^{-1}}$, whereas the widely studied FePt family does not enter this high-response regime even after compositional substitution. Here, a composition of the form $(X_{1-x}X^{'}_{x})(Y_{1-y}Y^{'}_{y})$ is classified as $XY$-based when $x,y \le 0.3$, such that $X$ and $Y$ remain the dominant constituent elements, while $X^{'}$ and $Y^{'}$ denote neighboring elements of $X$ and $Y$ in the periodic table.

Most compositions occupy the white region [$|\sigma_{xy}| \le 1500\,\mathrm{S\,cm^{-1}}$ and $|\alpha_{xy}| \le 4\,\mathrm{A\,m^{-1}\,K^{-1}}$], including the parent compounds NiPt and CoIr. A larger number of compositions satisfy $|\alpha_{xy}| \ge 4\,\mathrm{A\,m^{-1}\,K^{-1}}$ (red region) than $|\sigma_{xy}| \ge 1500\,\mathrm{S\,cm^{-1}}$ (blue region), indicating that the present compositional space contains more candidates with large ANC than with large AHC according to the adopted criteria. Only a few candidates occupy the green region [$|\sigma_{xy}| \ge 1500\,\mathrm{S\,cm^{-1}}$ and $|\alpha_{xy}| \ge 4\,\mathrm{A\,m^{-1}\,K^{-1}}$], indicating that simultaneously realizing large $|\sigma_{xy}|$ and $|\alpha_{xy}|$ is relatively uncommon within the investigated compositional space. Within the Mott relation, $\alpha_{xy}$ is governed by the energy variation of $\sigma_{xy}$ within the thermal window around the Fermi level. Consequently, a large $\alpha_{xy}$ favors a pronounced energy dependence of $\sigma_{xy}$ near $E_\mathrm{F}$, whereas $\alpha_{xy}$ tends to become small when $E_\mathrm{F}$ lies near an extremum of $\sigma_{xy}(E)$. The compositions exceeding the adopted criteria are predominantly chemically substituted compounds (open symbols), showing that chemical substitution substantially expands the number of candidates with large transverse responses beyond those accessible among the stoichiometric parent compounds alone. Nevertheless, several parent compounds, including NiPd, CoPt, CoRh, and GeMn, also exceed at least one of the adopted thresholds.

The candidates exhibiting the largest $|\sigma_{xy}|$ and $|\alpha_{xy}|$ are summarized in Table~\ref{tab1} and Table~\ref{tab2}, respectively. To represent a broad range of compositional families, only the composition with the largest $|\sigma_{xy}|$ or $|\alpha_{xy}|$ derived from each stoichiometric parent compound is included. Ranked according to $|\sigma_{xy}|$, the three leading compositions have $\sigma_{xy}$ values of \ce{(Co_{0.8}Fe_{0.2})(Ir_{0.7}Pt_{0.3})} [$2809\,\mathrm{S\,cm^{-1}}$], \ce{(Ni_{0.9}Co_{0.1})Pd} [$-2285\,\mathrm{S\,cm^{-1}}$], and \ce{(Ni_{0.9}Cu_{0.1})(Pt_{0.7}Au_{0.3})} [$-2261\,\mathrm{S\,cm^{-1}}$]. Among the highest-ranked AHC candidates, only GeMn remains an unsubstituted parent compound, further illustrating the effectiveness of chemical substitution in enhancing $\sigma_{xy}$ relative to the corresponding parent systems.

Only a few of the leading AHC candidates exhibit magnetization values larger than $3\,\mu_\mathrm{B}/\mathrm{f.u.}$, and these are mainly derived from GeMn-, MnSn-, and FePt-based systems. Even the FePt-derived composition \ce{(Fe_{0.9}Mn_{0.1})(Pt_{0.7}Ir_{0.3})} yields only $|\sigma_{xy}| = 1350\,\mathrm{S\,cm^{-1}}$. For many of the leading candidates, the small difference between $\sigma_{xy}$ and $\sigma^{\mathrm{max}}_{xy}$, together with the small value of $\Delta\epsilon_{\sigma}$, indicates that chemical substitution places the Fermi-level response close to the largest value accessible within the investigated energy window.

Ranked according to $|\alpha_{xy}|$, the three leading compositions have $\alpha_{xy}$ values of \ce{(Ni_{0.8}Co_{0.2})(Pt_{0.7}Ir_{0.3})} [$7.72\,\mathrm{A\,m^{-1}\,K^{-1}}$], \ce{Ni(Rh_{0.8}Pd_{0.2})} [$7.63\,\mathrm{A\,m^{-1}\,K^{-1}}$], and \ce{(Co_{0.8}Fe_{0.2})(Ir_{0.7}Os_{0.3})} [$7.18\,\mathrm{A\,m^{-1}\,K^{-1}}$]. Consistent with the distribution in Fig.~\ref{fig1}, CoIr- and NiPt-based alloys exhibit particularly large responses even though the corresponding parent compounds themselves show relatively small $|\sigma_{xy}|$ and $|\alpha_{xy}|$. With the exception of GeMn-based compositions, the leading ANC candidates exhibit magnetization values of $m \le 2.1\,\mu_\mathrm{B}/\mathrm{f.u.}$. This tendency motivates a more systematic examination of the relationship between magnetization, magnetic configuration, and anomalous transport in the following section. Similarly, for many of the leading ANC candidates, the small difference between $\alpha_{xy}$ and $\alpha^{\mathrm{max}}_{xy}$, together with the small value of $\Delta\epsilon_{\alpha}$, indicates that chemical substitution places the Fermi-level ANC close to the largest value accessible within the investigated energy window.

\subsection{Distribution of AHC and ANC with magnetization}\label{subsec3}

\begin{figure}[p] % Do NOT use \begin{figure*}
	\centering
	\includegraphics[width=\textwidth]{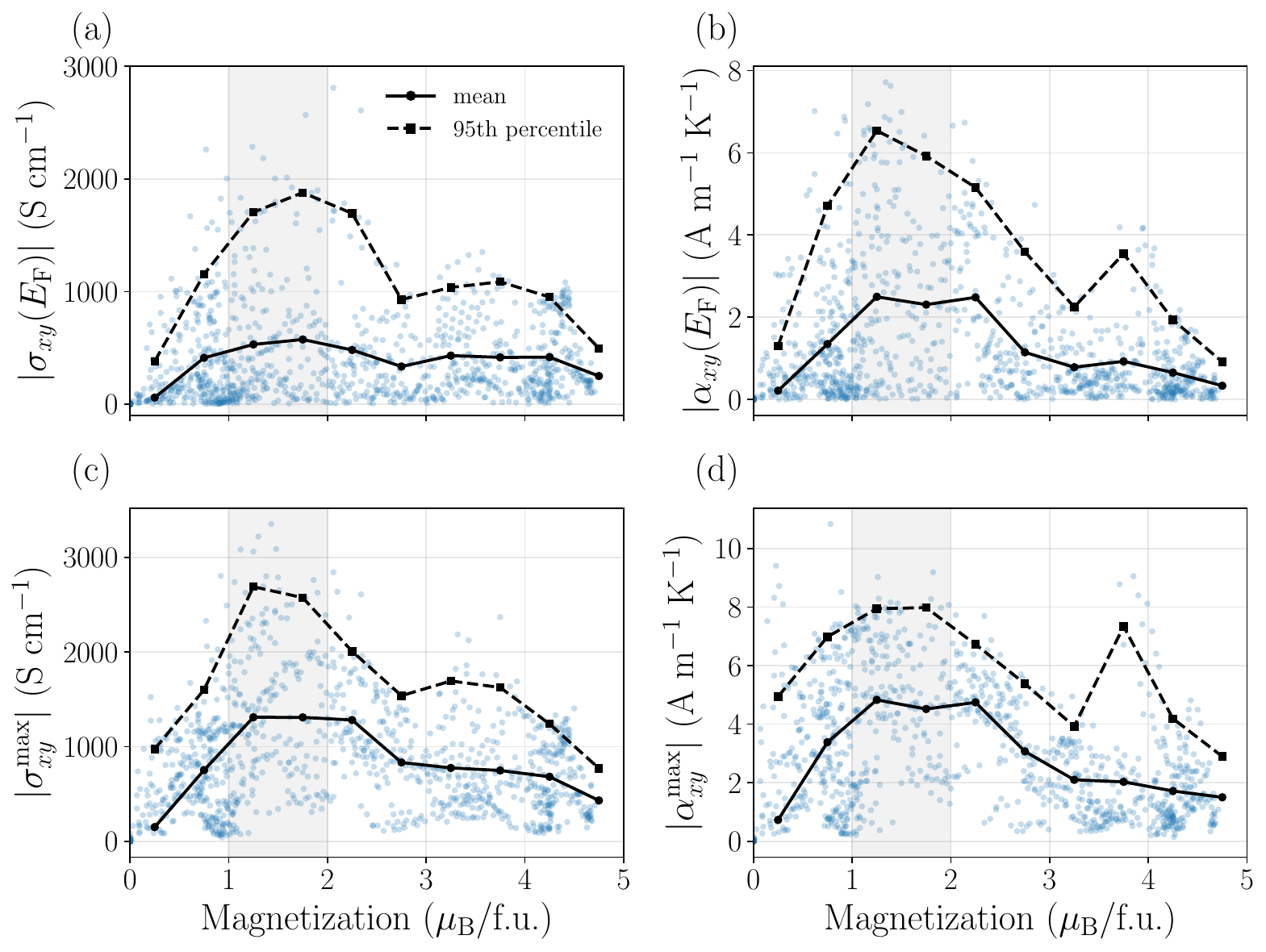} % for an image file named example_figure.*
\caption{\textbf{Statistics of $|\sigma_{xy}|$ and $|\alpha_{xy}|$ as a function of magnetization.}
Distribution of the absolute values of (\textbf{a}) $\sigma_{xy}$ at $T=0\,\mathrm{K}$, (\textbf{b}) $\alpha_{xy}$ at $T=300\,\mathrm{K}$, (\textbf{c}) $\sigma^{\mathrm{max}}_{xy}$ at $T=0\,\mathrm{K}$, and (\textbf{d}) $\alpha^{\mathrm{max}}_{xy}$ at $T=300\,\mathrm{K}$ as a function of magnetization for the ferromagnetic subset of the investigated alloys. The solid and dashed curves represent the mean and 95th percentile within magnetization bins of width $0.5\,\mu_\mathrm{B}/\mathrm{f.u.}$, respectively. The gray shaded region highlights the magnetization range of $1-2\,\mu_\mathrm{B}/\mathrm{f.u.}$, where the largest mean and 95th-percentile responses are observed.
}
	\label{fig2} % give each figure a logical label name
\end{figure}

To characterize the average and upper envelope of the response distribution, we calculate the mean and 95th percentile of $|\sigma_{xy}|$ and $|\alpha_{xy}|$ within magnetization bins of width $0.5\,\mu_\mathrm{B}/\mathrm{f.u.}$, shown by the solid and dashed curves in Fig.~\ref{fig2}, respectively. Here, we restrict the analysis to the ferromagnetic subset, for which magnetization provides a more meaningful measure of exchange splitting than in ferrimagnets, where a small net magnetization can instead arise from cancellation between antiparallel local moments. The corresponding analysis for the ferrimagnetic subset is provided in the Supplementary Information (Supplementary Note 1). For each magnetization bin, 95\% of the compositions lie below the 95th-percentile value, while the upper 5\% exceed it, providing a measure of the largest responses that is less sensitive to individual outliers. Both $|\sigma_{xy}(E_\mathrm{F})|$ and $|\alpha_{xy}(E_\mathrm{F})|$ exhibit their largest 95th-percentile values and mean values in the lower-to-intermediate magnetization range of $1-2\,\mu_\mathrm{B}/\mathrm{f.u.}$ [Fig.~\ref{fig2}(a)-(b)]. The corresponding maximum values within the calculated energy range, $|\sigma^{\mathrm{max}}_{xy}|$ and $|\alpha^{\mathrm{max}}_{xy}|$, follow similar trends [Fig.~\ref{fig2}(c)-(d)]. The corresponding mean-value curves are broader and less sharply peaked, likely because of averaging over a wider distribution of responses within each magnetization bin. Additional subpeaks appear in the range of $3-4\,\mu_\mathrm{B}/\mathrm{f.u.}$ and originate mainly from FePt-based compositions, consistent with the candidates listed in Table~\ref{tab1} and Table~\ref{tab2}. These results show that large AHC and ANC are not restricted to compounds with the largest magnetization values and instead preferentially emerge within specific lower-to-intermediate magnetization ranges.

\begin{figure}[p] % Do NOT use \begin{figure*}
	\centering
	\includegraphics[width=\textwidth]{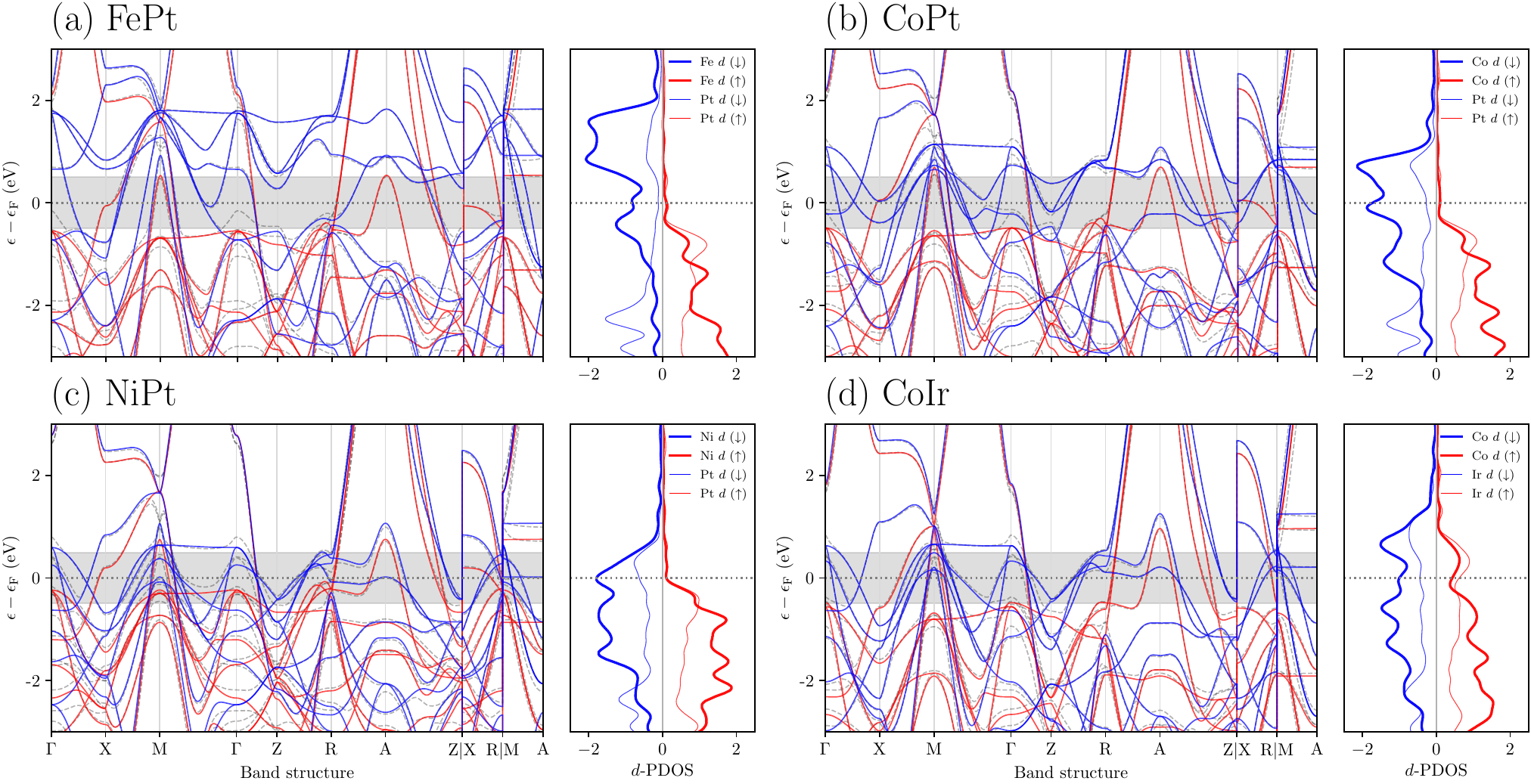} % for an image file named example_figure.*
\caption{\textbf{Band structures and density of states of representative parent compounds.}
Band structures along high-symmetry paths and atom- and spin-resolved densities of states for (\textbf{a}) FePt, (\textbf{b}) CoPt, (\textbf{c}) NiPt, and (\textbf{d}) CoIr. Red and blue curves represent majority- and minority-spin states, respectively. Gray dashed curves show the band structures calculated with spin-orbit coupling. The gray shaded region denotes the energy range from $-0.5$ to $0.5\,\mathrm{eV}$ relative to the Fermi level, which is indicated by the horizontal gray dotted line.
}
	\label{fig3} % give each figure a logical label name
\end{figure}

To investigate why large $|\sigma_{xy}|$ and $|\alpha_{xy}|$ preferentially emerge from particular ferromagnetic parent compounds in this magnetization regime, Fig.~\ref{fig3} compares the band structures and densities of states (DOSs) of four representative ferromagnetic parent compounds: FePt, CoPt, NiPt, and CoIr. In a ferromagnet, the exchange interaction lowers the energy of one spin channel relative to the other, displacing the majority- and minority-spin bands in energy. This displacement is the exchange splitting: it governs how unequally the two spin channels are occupied and therefore sets the magnetization, which is why the net magnetization serves as a practical measure of the exchange splitting in the screening above. Across the sequence from FePt to CoIr, the exchange splitting generally decreases, while the availability of \textit{d} states from both spin channels near the Fermi level increases. The well-known $L1_0$ compounds FePt and CoPt exhibit relatively large magnetic moments and exchange splitting, resulting in a pronounced energy separation between the majority- and minority-spin DOSs [Fig.~\ref{fig3}(a)-(b)]. In FePt, the dominant Fe \textit{d} minority-spin DOS peak lies above the Fermi level, although part of the minority-spin DOS remains near it. Replacing Fe with Co introduces additional electrons and shifts the Co \textit{d} minority-spin states in CoPt closer to the Fermi level than the corresponding Fe \textit{d} states in FePt.

Further electron filling in NiPt places the Ni \textit{d} and Pt \textit{d} minority-spin states near the Fermi level, while the majority-spin states remain slightly below it, consistent with a reduced exchange splitting. CoIr exhibits a still smaller magnetic moment and exchange splitting, with \textit{d} states from both spin channels and both atomic sites distributed broadly around the Fermi level. The resulting coexistence of majority- and minority-spin bands near the Fermi level provides more opportunities for chemical substitution to modify band crossings and other near-degenerate electronic states. This trend helps explain why NiPt- and CoIr-based ferromagnetic parent compounds provide particularly favorable starting points for achieving large Berry-curvature-driven transport responses upon chemical substitution, although the magnitude of the final response also depends on the detailed orbital character and spin--orbit-induced modification of the bands.

\section{Discussion}\label{sec3}

Our results indicate that the magnitude and chemical tunability of Berry-curvature-driven transport are strongly influenced by the electronic structure inherited from the parent compound. In particular, the ferromagnetic parent compounds NiPt and CoIr exhibit relatively small magnetic moments and reduced exchange splitting, allowing bands from both spin channels to remain available near the Fermi level. Together with the strong spin--orbit coupling provided by heavy elements such as Pt and Ir, this band configuration creates more opportunities for chemical substitution to modify near-Fermi-level crossings and generate electronic structures associated with large Berry curvature. This trend is consistent with previous high-throughput calculations of the anomalous Hall effect, which found that large AHC can emerge in compounds with relatively small magnetization, particularly in systems containing heavy elements~\cite{zelezny2023}. The present results further show that the parent compound strongly influences how effectively these anomalous transport responses can be enhanced through compositional tuning.

\begin{figure}[p] % Do NOT use \begin{figure*}
	\centering
	\includegraphics[width=\textwidth]{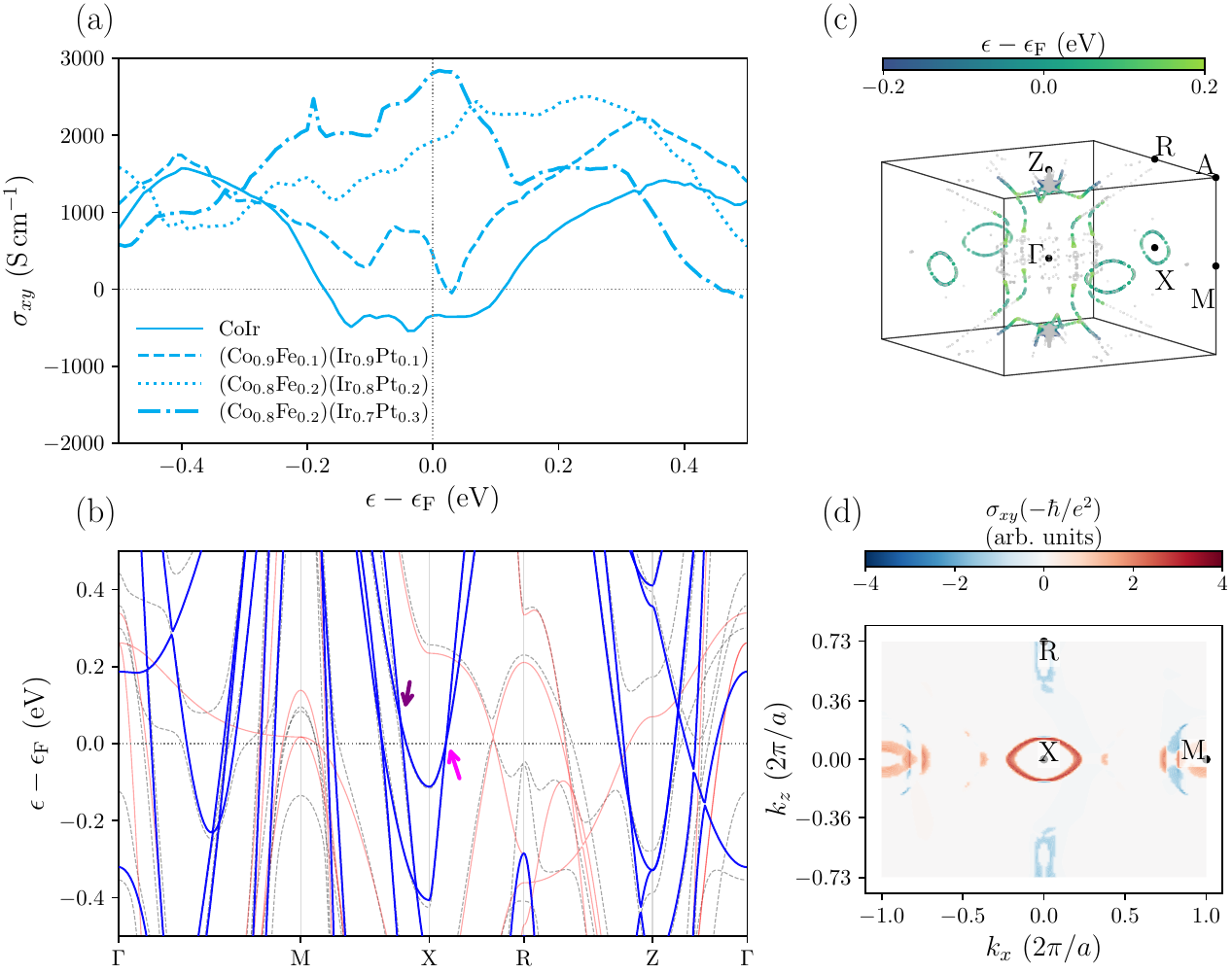} % for an image file named example_figure.*
\caption{\textbf{Electronic origin of the enhanced $\sigma_{xy}$ in CoIr-derived alloys.}
(\textbf{a}) Energy dependence of $\sigma_{xy}$ for CoIr, \ce{(Co_{0.9}Fe_{0.1})(Ir_{0.9}Pt_{0.1})}, \ce{(Co_{0.8}Fe_{0.2})(Ir_{0.8}Pt_{0.2})}, and \ce{(Co_{0.8}Fe_{0.2})(Ir_{0.7}Pt_{0.3})}. (\textbf{b}) Band structure of \ce{(Co_{0.8}Fe_{0.2})(Ir_{0.7}Pt_{0.3})}. Red and blue curves represent majority- and minority-spin bands, respectively, while black dashed curves show the bands calculated with spin-orbit coupling. The purple and magenta arrows mark the minority-spin band crossing near the Fermi level associated with the nodal-line network. (\textbf{c}) Symmetry-related nodal lines formed by the two minority-spin bands on the Brillouin-zone faces containing the X--R and X--M paths. (\textbf{d}) $k$-decomposed $\sigma_{xy}$ on the plane containing the X--R and X--M paths. The regions contributing strongly to $\sigma_{xy}$ coincide with the locations of the spin-orbit-gapped nodal lines.
}
	\label{fig4} % give each figure a logical label name
\end{figure}

The enhancement of AHC from CoIr [$-340\,\mathrm{S\,cm^{-1}}$] to \ce{(Co_{0.8}Fe_{0.2})(Ir_{0.7}Pt_{0.3})} [$2809\,\mathrm{S\,cm^{-1}}$] is particularly noteworthy. Figure~\ref{fig4}(a) shows how the energy dependence of $\sigma_{xy}$ evolves from the CoIr parent compound to \ce{(Co_{0.9}Fe_{0.1})(Ir_{0.9}Pt_{0.1})}, and then to \ce{(Co_{0.8}Fe_{0.2})(Ir_{0.8}Pt_{0.2})}, before reaching its maximum value at the Fermi level in \ce{(Co_{0.8}Fe_{0.2})(Ir_{0.7}Pt_{0.3})}. The shape of the energy-dependent $\sigma_{xy}$ changes substantially upon chemical substitution, indicating that the enhancement cannot be described simply as a rigid shift of the Fermi level within the electronic structure of CoIr.

The band crossing associated with this large AHC is located along the X--R path in the Brillouin zone, where two minority-spin bands cross near the Fermi level [magenta arrow, Fig.~\ref{fig4}(b)]. The same two bands also cross along the X--M path [purple arrow, Fig.~\ref{fig4}(b)], forming symmetry-related nodal lines on the Brillouin-zone faces [Fig.~\ref{fig4}(c)]. Importantly, the corresponding nodal-line structure is absent in the CoIr parent compound, as shown in the Supplementary Information (Supplementary Note 2). This comparison demonstrates that chemical substitution modifies the electronic structure beyond a simple rigid-band shift and that the VCA captures the resulting reconstruction of the relevant bands. For magnetization along [001], spin--orbit coupling gaps the nodal lines on the four equivalent Brillouin-zone faces containing the magnetization axis, generating large Berry curvature and thereby contributing strongly to $\sigma_{xy}$. This correspondence is evident in the $k$-decomposed $\sigma_{xy}$ on the plane containing the X--R and X--M paths in Fig.~\ref{fig4}(d), where the regions contributing strongly to $\sigma_{xy}$ closely follow the locations of the spin--orbit-gapped nodal lines. When the magnetization is rotated to [100] or [010], only the two nodal lines lying on Brillouin-zone faces containing the corresponding magnetization axis are gapped, whereas those on the perpendicular faces remain largely ungapped, consistent with mirror-symmetry protection. Consequently, $\sigma_{yz}$ for magnetization along [100] and $\sigma_{xz}$ for magnetization along [010] reach approximately $1545\,\mathrm{S\,cm^{-1}}$ at the Fermi level, roughly half of the $\sigma_{xy}$ value obtained for magnetization along [001], as discussed in the Supplementary Information (Supplementary Note 3 and 4).

\begin{figure}[p] % Do NOT use \begin{figure*}
	\centering
	\includegraphics[width=\textwidth]{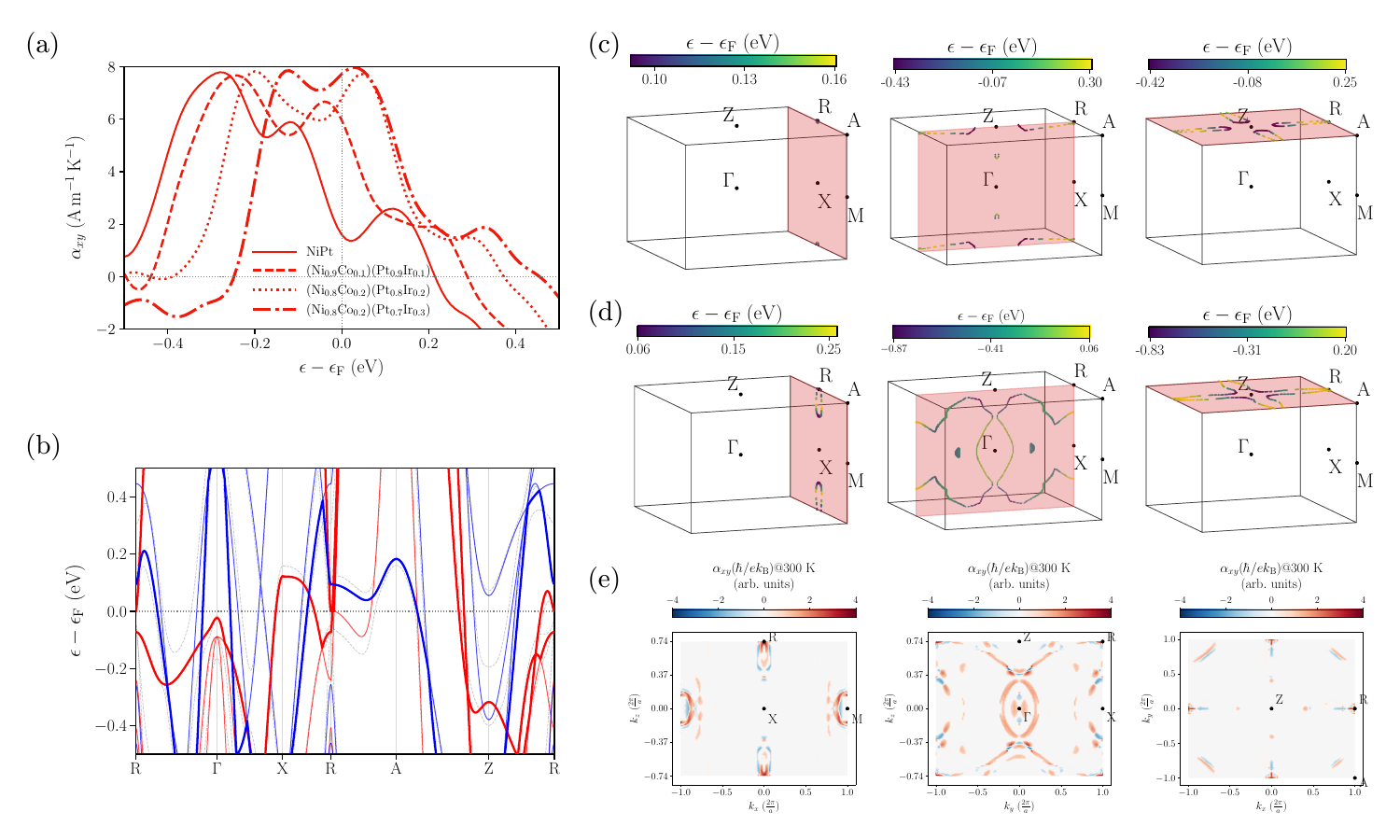} % for an image file named example_figure.*
\caption{\textbf{Electronic origin of the enhanced $\alpha_{xy}$ in NiPt-derived alloys.}
(\textbf{a}) Energy dependence of $\alpha_{xy}$ for NiPt, \ce{(Ni_{0.9}Co_{0.1})(Pt_{0.9}Ir_{0.1})}, \ce{(Ni_{0.8}Co_{0.2})(Pt_{0.8}Ir_{0.2})}, and \ce{(Ni_{0.8}Co_{0.2})(Pt_{0.7}Ir_{0.3})}. (\textbf{b}) Band structure of \ce{(Ni_{0.8}Co_{0.2})(Pt_{0.7}Ir_{0.3})}. Red and blue curves represent majority- and minority-spin bands, respectively, while black dashed curves show the bands calculated with spin-orbit coupling. The bold curves highlight the upper and lower majority-spin bands and the minority-spin band associated with the crossing structures discussed in (\textbf{c}) and (\textbf{d}). (\textbf{c}) Small closed crossing contours near the R point formed by the upper majority-spin and minority-spin bands highlighted in (\textbf{b}). (\textbf{d}) More extended nodal-line structure formed by the lower majority-spin and minority-spin bands highlighted in (\textbf{b}). (\textbf{e}) Corresponding $k$-decomposed $\alpha_{xy}$. In (\textbf{c})-(\textbf{e}), the left, middle, and right panels show planes containing the X--R and X--M paths, the $\Gamma$--Z and $\Gamma$--X paths, and the Z--R and R--A paths, respectively. The localized regions contributing strongly to $\alpha_{xy}$ in (\textbf{e}) closely coincide with the near-R crossing contours in (\textbf{c}), while additional features show correspondence with the more extended nodal-line structure in (\textbf{d}).
}
	\label{fig5} % give each figure a logical label name
\end{figure}

The ANC at the Fermi level increases from $1.59\,\mathrm{A\,m^{-1}\,K^{-1}}$ in NiPt to progressively larger values in \ce{(Ni_{0.9}Co_{0.1})(Pt_{0.9}Ir_{0.1})} and \ce{(Ni_{0.8}Co_{0.2})(Pt_{0.8}Ir_{0.2})}, reaching $7.72\,\mathrm{A\,m^{-1}\,K^{-1}}$ in \ce{(Ni_{0.8}Co_{0.2})(Pt_{0.7}Ir_{0.3})}, as shown in Fig.~\ref{fig5}(a). In contrast to the substantial evolution of $\sigma_{xy}(E)$ observed in the CoIr-derived series, the overall shape of $\alpha_{xy}(E)$ in the NiPt-derived series remains comparatively similar while shifting toward lower energy with increasing substitution. This behavior indicates that the electronic structure relevant to the ANC undergoes less pronounced reconstruction than in the CoIr-derived case and that the enhancement is associated primarily with bringing favorable electronic features closer to the Fermi level.

The electronic states most relevant to this enhancement are concentrated near the R point, where several majority-spin bands lie close in energy to a minority-spin band near the Fermi level [Fig.~\ref{fig5}(b)]. Among these states, two majority-spin bands, hereafter referred to as the upper and lower majority-spin bands, together with a nearby minority-spin band, form distinct crossing structures in the Brillouin zone. The corresponding bands are highlighted by the bold curves in Fig.~\ref{fig5}(b). The crossing between the upper majority-spin band and the minority-spin band produces small closed contours localized near the R point [Fig.~\ref{fig5}(c)], whereas the crossing between the lower majority-spin band and the minority-spin band forms a more extended nodal-line structure [Fig.~\ref{fig5}(d)].

Figures~\ref{fig5}(c)-(e) compare these crossing structures with the corresponding $k$-decomposed $\alpha_{xy}$. On the plane passing through the R point shown in the left panels, the small closed crossing contours formed by the upper majority-spin and minority-spin bands [Fig.~\ref{fig5}(c)] closely coincide with localized regions that contribute strongly to $\alpha_{xy}$ [Fig.~\ref{fig5}(e)]. Related features are also visible on the central plane and on the top Brillouin-zone face shown in the middle and right panels, respectively. This momentum-space correspondence indicates that the near-R crossing contours formed by the upper majority-spin and minority-spin bands contribute substantially to the large ANC. The more extended nodal-line structure formed by the lower majority-spin and minority-spin bands [Fig.~\ref{fig5}(d)] also shows a correspondence with features in the $k$-decomposed $\alpha_{xy}$ [Fig.~\ref{fig5}(e)], indicating an additional contribution. However, the associated $\alpha_{xy}$ features are weaker than those corresponding to the small closed contours near R, suggesting that the latter provide the dominant contribution among the crossing structures identified here.

Because both majority- and minority-spin bands participate in these near-degenerate states, crossings between bands of different spin character play an important role in the enhanced ANC. Contributions from crossings between nearby majority-spin bands may also be present, and the present analysis does not exclude their involvement. Comparison with the band structure of the NiPt parent compound in the Supplementary Information (Supplementary Note 5) further shows that the majority- and minority-spin band segments involved in these crossings retain broadly similar dispersions upon chemical substitution, while their relative energies shift. This behavior contrasts with the stronger band reconstruction found in the CoIr-derived AHC case.

The large ANC is further accompanied by a pronounced energy dependence of the anomalous Hall conductivity near the Fermi level, consistent with the relation between $\alpha_{xy}$ and the energy variation of $\sigma_{xy}$ within the thermal window, as discussed in the Supplementary Information (Supplementary Note 6). These results illustrate how reduced exchange splitting can promote the coexistence of bands from both spin channels near the Fermi level. Chemical substitution can then shift, reshape, or generate relevant band crossings, producing a strong energy variation of the Berry-curvature-driven transverse response that is favorable for large ANC. At the same time, the increased availability of bands near the Fermi level also raises the likelihood of relevant crossings between bands with the same spin character. The CoIr- and NiPt-derived systems therefore represent complementary examples in which reduced exchange splitting enhances chemical tunability without imposing a single universal microscopic type of band crossing.

Taken together, our results demonstrate that giant AHC and ANC can be realized through chemical substitution when the starting ferromagnetic parent compound provides a favorable near-Fermi-level electronic structure. Ferromagnetic parent compounds with relatively small exchange splitting maintain bands from both spin channels near the Fermi level, making their band crossings and Berry-curvature-driven responses particularly sensitive to compositional modification. Exchange splitting therefore provides a useful descriptor for assessing the chemical tunability of ferromagnetic parent compounds rather than merely ranking their transport responses in the stoichiometric form. This parent-compound selection principle moves beyond brute-force high-throughput screening by identifying compositional families in which chemical substitution is more likely to produce large anomalous transport responses. It can consequently guide both computational exploration and combinatorial experiments toward more promising regions of materials space. A natural next step is to combine this descriptor with large-scale stability screening, for which extensive databases of formation energies, phonon stability, and magnetic critical temperatures are now becoming available~\cite{xiao2025,xiao2026}. Such a combination would be particularly valuable here, since parent compounds with small exchange splitting also tend to exhibit lower Curie temperatures, so that chemical tunability and thermal robustness must ultimately be optimized together. The predicted CoIr- and NiPt-derived ferromagnetic alloys provide specific candidates for experimental validation and illustrate the potential of exchange-splitting-informed compositional design for thermoelectric waste-heat harvesting and AHE-based magnetic-recording technologies.

\section{Methods}\label{sec4}

Transport properties were computed automatically using an in-house-developed Python workflow interfaced with {\scriptsize QUANTUM ESPRESSO} (QE)~\cite{giannozzi2009,giannozzi2017}, {\scriptsize WANNIER90}~\cite{marzari1997,souza2001}, and {\scriptsize WANNIERTOOLS}~\cite{wu2018}. The initial stoichiometric $L1_0$ compounds were selected from the Open Quantum Materials Database (OQMD)~\cite{kirklin2015}. Compounds with an energy above the convex hull of no more than $0.3\,\mathrm{eV}$ and a finite magnetic moment larger than $0.1\,\mu_\mathrm{B}$ per formula unit were retained, yielding 67 parent compounds. The primitive tetragonal $L1_0$ unit cell containing one formula unit of $XY$ was used throughout. The lattice parameters and atomic positions of each substituted composition were relaxed using scalar-relativistic pseudopotentials before the fully relativistic electronic-structure and transport calculations.

Chemical substitution was simulated using the virtual crystal approximation (VCA)~\cite{bellaiche2000} for compositions of the form $(X_{1-x}X^{'}_{x})(Y_{1-y}Y^{'}_{y})$, where $X^{'}$ ($Y^{'}$) denotes an element neighboring $X$ ($Y$) on either side in the periodic table, and $x,y = 0,0.1,0.2,0.3$. The VCA pseudopotential for each substituted site was constructed by concentration-weighted interpolation of the corresponding elemental pseudopotentials. This procedure yielded transport data for 2251 chemically substituted candidates.

All candidates were treated within collinear ferromagnetic or ferrimagnetic configurations, with the initial magnetization aligned along the [001] crystal axis. For each composition, the considered collinear magnetic configurations were compared, and the configuration with the lowest total energy was used for the subsequent transport calculations. Self-consistent electronic-structure calculations were performed using {\scriptsize QUANTUM ESPRESSO}~\cite{giannozzi2009,giannozzi2017} within the generalized gradient approximation of Perdew, Burke, and Ernzerhof (PBE)~\cite{perdew1996}. Fully relativistic optimized norm-conserving Vanderbilt (ONCV) pseudopotentials~\cite{hamann2013} obtained from PseudoDojo~\cite{setten2018} were employed. Spin-orbit coupling was included in all calculations used to evaluate the transport properties and Berry curvature. Kinetic-energy cutoffs of 100 and 400 Ry were used for the wave functions and charge density, respectively. For the self-consistent calculations, the Brillouin zone was sampled using a $k$-point mesh with a reciprocal-space density of approximately $450\,\mathrm{\AA^3}$, defined as the total number of $k$ points divided by the Brillouin-zone volume. Electronic occupations were treated using Marzari--Vanderbilt--DeVita--Payne cold smearing~\cite{marzari1999} with a smearing width of $0.015\,\mathrm{Ry}$. The self-consistent-field convergence threshold was set to $10^{-10}\,\mathrm{Ry}$.

Maximally localized Wannier functions were constructed using {\scriptsize WANNIER90}~\cite{marzari1997,souza2001}. The initial localized subspace was generated using the selected columns of the density matrix (SCDM) method~\cite{vitale2020,damle2015,damle2018}, which is well suited to automated high-throughput Wannierization. The outer and frozen disentanglement windows were adjusted automatically within the workflow for each composition. The quality of the Wannier interpolation was verified by comparing the interpolated and first-principles band structures within $1\,\mathrm{eV}$ of the Fermi level, corresponding to the energy range relevant to the transport calculations.

The anomalous Hall conductivity $\sigma_{xy}$ was evaluated from the Brillouin-zone integral of the Berry curvature using the Wannier-interpolated Hamiltonian. The transport calculations employed a dense $k$-point mesh containing 12 times more points along each reciprocal-lattice direction than the mesh used for the self-consistent calculation. The anomalous Nernst conductivity $\alpha_{xy}$ was calculated at $300\,\mathrm{K}$ from the energy-dependent anomalous Hall conductivity using the finite-temperature Mott relation described by Xiao \textit{et al.}~\cite{xiao2006}. Energy-dependent values of $\sigma_{xy}$ and $\alpha_{xy}$ were evaluated over the range $[-1,1]\,\mathrm{eV}$ relative to the Fermi level using an energy step of $0.01\,\mathrm{eV}$.

To characterize how the computed responses are distributed with magnetization, the candidates were grouped into bins of net magnetization $|M|$ of width $0.5\,\mu_\mathrm{B}$/f.u., each covering the half-open interval $[M_i,\,M_i+0.5)$ and plotted at its midpoint. Within each bin we evaluated the mean and the 95th percentile of $|\sigma_{xy}(E_\mathrm{F})|$, $|\alpha_{xy}(E_\mathrm{F})|$, $|\sigma^{\mathrm{max}}_{xy}|$, and $|\alpha^{\mathrm{max}}_{xy}|$, where the latter two denote the largest absolute values within the scanned window of $[-0.3,0.3]\,\mathrm{eV}$ relative to the Fermi level; percentiles were obtained by linear interpolation between the two nearest order statistics. Bins containing fewer than 10 compositions were excluded from the binned curves, while all individual compositions are shown as scatter points. Identical bin edges were used for the ferromagnetic and ferrimagnetic subsets.

The Berry curvature, $k$-resolved contributions to $\sigma_{xy}$ and $\alpha_{xy}$, and nodal-line structures of selected candidates were analyzed using {\scriptsize WANNIERTOOLS}~\cite{wu2018}. Nodal lines were identified from crossings in the spin-resolved band structures calculated without spin-orbit coupling, whereas their spin-orbit-induced gapping and associated Berry-curvature distributions were examined using the fully relativistic Wannier Hamiltonian. Calculations with magnetization along [100], [010], and [001] were additionally performed for selected candidates to analyze the anisotropy of the anomalous transport response. Further details of the automated workflow follow the procedure described in Ref.~\cite{xing2024}.

\section*{Acknowledgements}
We thank Y. Iwasaki, T. Nakatani, Y. Sakuraba, T. Tadano, and E. Xiao of NIMS for helpful discussions. Theoretical calculations in this work were performed using the Numerical Materials Simulator at NIMS. This work was partly supported by JST-CREST (Grant No. JPMJCR21O1), JSPS KAKENHI (Grant Nos. JP22H04966, JP24K00932, JP25H00743, and JP26K00993), the MEXT Program for Data Creation and Utilization-Type Material Research and Development Project (Digital Transformation Initiative Center for Magnetic Materials; Grant No. JPMXP1122715503), and the MEXT Initiative to Establish Next-generation Novel Integrated Circuits Centers (X-NICS; Grant No. JPJ011438).
\clearpage

\clearpage

\clearpage

\clearpage

\clearpage

\clearpage

\begin{table}[p] % Do NOT use \begin{table*}
	\centering
\caption{\textbf{Top $L1_0$ alloy candidates with large $|\sigma_{xy}|$ at the Fermi level.}
For each stoichiometric parent compound, only the derived composition with the largest $|\sigma_{xy}|$ is listed. The magnetization $m$ is given per formula unit. $\sigma_{xy}$ and $\alpha_{xy}$ denote the anomalous Hall conductivity at $T=0\,\mathrm{K}$ and the anomalous Nernst conductivity at $T=300\,\mathrm{K}$, respectively, evaluated at the Fermi level. $\sigma^{\mathrm{max}}_{xy}$ and $\alpha^{\mathrm{max}}_{xy}$ denote the corresponding maximum values within the energy range $[-0.3,0.3]\,\mathrm{eV}$ relative to the Fermi level. $\Delta\epsilon_{\sigma}$ and $\Delta\epsilon_{\alpha}$ denote the energy shifts of $\sigma^{\mathrm{max}}_{xy}$ and $\alpha^{\mathrm{max}}_{xy}$, respectively, relative to the Fermi level.
}
	\label{tab1} % give each table a logical label name
	
	\begin{tabular}{l c r r r r r r} % four columns, alignment for each
		\\
		\hline
%		Sample & $A$ & $B$ & $C$\\
%		 & (unit) & (unit) & (unit)\\
      Candidates & $m$ & $\sigma_{xy}$ & $\alpha_{xy}$ & $\sigma^{\mathrm{max}}_{xy}$ & $\Delta\epsilon_{\sigma}$ & $\alpha^{\mathrm{max}}_{xy}$ & $\Delta\epsilon_{\alpha}$ \\
       & $(\mu_\mathrm{B}/\mathrm{f.u.})$ & $(\mathrm{S\,cm^{-1}})$ & $(\mathrm{A\,m^{-1}K^{-1}})$ & $(\mathrm{S\,cm^{-1}})$ & $(\mathrm{eV})$ & $(\mathrm{A\,m^{-1}K^{-1}})$ & $(\mathrm{eV})$ \\ 
		\hline
\ce{(Co_{0.8}Fe_{0.2})(Ir_{0.7}Pt_{0.3})} & 2.06 & 2809.07 & -0.18 & 2841.49 & 0.01 & 6.36 & 0.09 \\
\ce{(Ni_{0.9}Co_{0.1})Pd} & 1.24 & -2285.69 & -1.91 & -2353.32 & -0.01 & 6.36 & -0.14 \\
\ce{(Ni_{0.9}Cu_{0.1})(Pt_{0.7}Au_{0.3})} & 0.77 & -2261.69 & 1.08 & -2328.56 & 0.01 & 5.96 & -0.30 \\
\ce{(Co_{0.9}Ni_{0.1})(Rh_{0.9}Ru_{0.1})} & 1.84 & 1876.27 & -0.20 & 1958.52 & -0.04 & 7.32 & 0.23 \\
\ce{GeMn} & 3.09 & 1728.72 & -1.58 & 1738.85 & -0.01 & -6.29 & -0.06 \\
\ce{(Co_{0.9}Ni_{0.1})(Fe_{0.7}Co_{0.3})} & 0.44 & -1468.98 & 3.76 & -1786.22 & 0.04 & 7.56 & -0.16 \\
\ce{(Mn_{0.7}Fe_{0.3})(Sn_{0.9}In_{0.1})} & 3.33 & 1390.89 & -1.09 & 1390.89 & 0.00 & -4.49 & -0.05 \\
\ce{(Fe_{0.9}Mn_{0.1})(Pt_{0.7}Ir_{0.3})} & 3.57 & 1350.14 & 0.28 & 1764.60 & -0.25 & -2.26 & -0.30 \\
\ce{(Mn_{0.8}Cr_{0.2})(Os_{0.8}Re_{0.2})} & 0.99 & 1348.78 & 2.39 & 1396.06 & -0.02 & -7.04 & -0.12 \\
\ce{(Co_{0.7}Ni_{0.3})(Mn_{0.7}Fe_{0.3})} & 1.79 & 1203.81 & -0.33 & 1267.23 & -0.06 & -4.65 & -0.12 \\
		\hline
	\end{tabular}
\end{table}
\clearpage

\begin{table}[p] % Do NOT use \begin{table*}
	\centering
\caption{\textbf{Top $L1_0$ alloy candidates with large $|\alpha_{xy}|$ at the Fermi level.}
For each stoichiometric parent compound, only the derived composition with the largest $|\alpha_{xy}|$ is listed. The magnetization $m$ is given per formula unit. $\sigma_{xy}$ and $\alpha_{xy}$ denote the anomalous Hall conductivity at $T=0\,\mathrm{K}$ and the anomalous Nernst conductivity at $T=300\,\mathrm{K}$, respectively, evaluated at the Fermi level. $\sigma^{\mathrm{max}}{xy}$ and $\alpha^{\mathrm{max}}{xy}$ denote the corresponding maximum values within the energy range $[-0.3,0.3]\,\mathrm{eV}$ relative to the Fermi level. $\Delta\epsilon_{\sigma}$ and $\Delta\epsilon_{\alpha}$ denote the energy shifts of $\sigma^{\mathrm{max}}_{xy}$ and $\alpha^{\mathrm{max}}_{xy}$, respectively, relative to the Fermi level.
}
	\label{tab2} % give each table a logical label name
	
	\begin{tabular}{l c r r r r r r} % four columns, alignment for each
		\\
		\hline
      Candidates & $m$ & $\sigma_{xy}$ & $\alpha_{xy}$ & $\sigma^{max}_{xy}$ & $\Delta\epsilon_{\sigma}$ & $\alpha^{max}_{xy}$ & $\Delta\epsilon_{\alpha}$ \\
       & $(\mu_\mathrm{B}/\mathrm{f.u.})$ & $(\mathrm{S\,cm^{-1}})$ & $(\mathrm{A\,m^{-1}K^{-1}})$ & $(\mathrm{S\,cm^{-1}})$ & $(\mathrm{eV})$ & $(\mathrm{A\,m^{-1}K^{-1}})$ & $(\mathrm{eV})$ \\ 
		\hline
\ce{(Ni_{0.8}Co_{0.2})(Pt_{0.7}Ir_{0.3})} & 1.34 & 262.44 & 7.72 & 2401.73 & -0.22 & 7.97 & 0.03 \\
\ce{Ni(Rh_{0.8}Pd_{0.2})} & 1.39 & -484.87 & 7.63 & 1597.63 & -0.27 & 8.16 & -0.03 \\
\ce{(Co_{0.8}Fe_{0.2})(Ir_{0.7}Os_{0.3})} & 1.16 & 297.43 & 7.18 & 1456.44 & -0.20 & 7.25 & 0.01 \\
\ce{(Cu_{0.7}Ni_{0.3})(Rh_{0.7}Pd_{0.3})} & 0.62 & 256.85 & 7.01 & 1052.07 & -0.07 & 7.09 & 0.01 \\
\ce{(Co_{0.7}Ni_{0.3})(Rh_{0.8}Ru_{0.2})} & 1.33 & 451.54 & -6.87 & 1230.56 & 0.10 & -6.87 & 0.00 \\
\ce{(Cu_{0.8}Ni_{0.2})(Ni_{0.7}Co_{0.3})} & 0.93 & -408.83 & 6.77 & -876.23 & -0.25 & 6.94 & -0.01 \\
\ce{(Co_{0.8}Ni_{0.2})(Ni_{0.8}Cu_{0.2})} & 2.02 & -204.35 & 6.66 & -1237.61 & 0.14 & 6.76 & 0.01 \\
\ce{(Mn_{0.7}Cr_{0.3})(Os_{0.7}Re_{0.3})} & 0.98 & 600.48 & -6.62 & 1439.22 & 0.09 & -6.75 & -0.02 \\
\ce{(Fe_{0.8}Co_{0.2})(Se_{0.9}Br_{0.1})} & 2.06 & 1071.57 & 6.37 & 1386.70 & -0.30 & 7.13 & 0.02 \\
\ce{(Ge_{0.9}As_{0.1})(Mn_{0.9}Cr_{0.1})} & 3.10 & 1392.75 & -5.87 & 1748.88 & 0.03 & -6.58 & -0.02 \\
		\hline
	\end{tabular}
\end{table}
\clearpage

\end{document}